# Thermodynamic calculation of the B–C system at pressures to 24 GPa

Vladimir Z. Turkevich [a] and Vladimir L. Solozhenko [b,*]

[a] *Institute for Superhard Materials, National Academy of Sciences of Ukraine, Kiev, 04074 Ukraine*

[b] *LSPM–CNRS, Université Paris Nord, 93430 Villetaneuse, France*

The evolution of topology of the B–C phase diagram has been studied at pressures up to 24 GPa using models of phenomenological thermodynamics with interaction parameters derived from experimental data on phase equilibria at high pressures and high temperatures.

The B–C system that includes $B_4C$ boron carbide, the basis of dense impact-resistant ceramics, is well understood at ambient pressure [1]. However, data on the B–C phase diagram at high pressure are absent in the literature. In this study thermodynamic calculations of high-pressure phase equilibria in the system were performed for the first time, and the phase diagrams at 7.7, 15, and 24 GPa were constructed.

The calculations were carried out using the Thermo-Calc software [2]. Thermodynamic data on phases of the B–C system at ambient pressure were taken from [1]. The liquid phase was described using the subregular solution model [3], and solid phases – in the framework of the Compound Energy Formalism (CEF) [4]. Pressure dependencies of molar volumes were represented using the Murnaghan approximation [5]. Bulk moduli, their pressure derivatives, and thermal expansion coefficients for β-$B_{106}$, graphite, diamond, and liquid phase were taken from [6, 7], while the data on high-pressure phases, γ-$B_{28}$ [8] and t′-$B_{52}$ [9], – from [10]. We also used literature data on the thermal expansion [11] and compressibility [12] of $B_4C$ boron carbide.

The molar volume of the liquid phase was described by the equation

$$V_L = V_B x_B + V_C x_C + \Delta V x_B x_C ,$$

where $\Delta V = -5.8 \cdot 10^{-6} - 5.0 \cdot 10^{-6} (x_B - x_C)$ is the mixing volume defined by the optimization of thermodynamic data from the pressure dependencies of melting temperatures of $B_4C$ and $B_4C$–graphite eutectics [13].

The evolution of the phase diagram of the B–C system under pressure is shown in the figure. In addition to the quantitative changes of the diagram parameters (equilibria temperatures, limiting solubilities), a change of the diagram topology is observed i.e. the diamond thermodynamic stability region appears, the L + $B_4C$ ⇄ β-$B_{106}$ peritectic changes to the L + γ-$B_{28}$ ⇄ $B_4C$ peritectic, the γ-$B_{28}$ ⇄ t′-$B_{52}$ equilibrium line appears, and the congruent type of $B_4C$ melting changes to the incongruent one. The limiting solubility of boron in diamond at 2200 K and 24 GPa is 0.1 at % (figure *d*), therefore, diamond-like $BC_5$ [14] as well as graphite-like boron-substituted (up to 40 at % boron) carbons [15] are not the equilibrium but metastable solid solutions and thus are not presented in the equilibrium phase diagram of the B–C system.

* vladimir.solozhenko@univ-paris13.fr

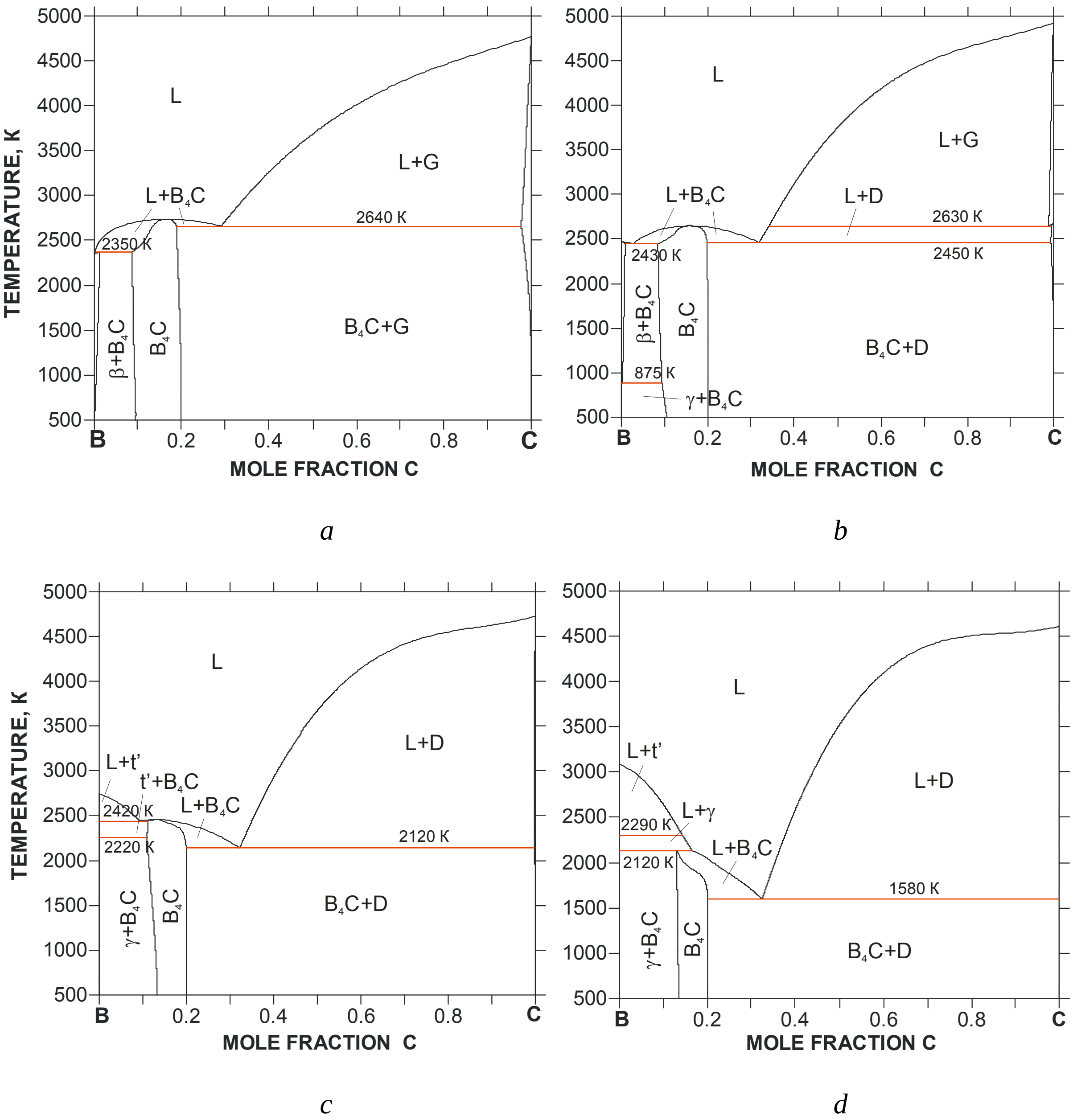

Phase diagram of the B–C system at 0.1 MPa (*a*), 7.7 GPa (*b*), 15 GPa (*c*) and 24 GPa (*d*); β, γ, t′ are boron allotropes (β-$B_{106}$, γ-$B_{28}$ and t′-$B_{52}$, respectively); L is the liquid phase of the B–C system; G is graphite; D is diamond.

V.L.S. is grateful to the Agence Nationale de la Recherche for financial support (grant ANR-2011-BS08-018).